\documentclass[a4paper,11pt]{article}
\usepackage{jcappub} 
\usepackage{lineno}

\title{A Dynamical Scalar Field Model for Dark Energy: Addressing the Hubble Tension and Cosmic Evolution}

\author[a,1]{Arpit Kottur\note{Corresponding author.}}
\author[a]{Jui Mahajan}
\author[a]{and Raka Dabhade}
\affiliation[a]{Department of Physics, Fergusson College (Autonomous),\\Pune, India}

\abstract{We propose a dynamical dark energy model based on a canonical scalar field with a hybrid potential of the form $V(\phi) = V_{0}e^{-\lambda\phi} + V_{1}\phi^{n}$. We constrain the model's 11-dimensional parameter space using a comprehensive combination of cosmological data, including the Planck 2018 Cosmic Microwave Background (CMB) power spectra, Baryon Acoustic Oscillations (BAO), the Pantheon+ supernova sample, SH0ES and the matter power spectrum from SDSS. The model provides an excellent fit to the data, with a reduced chi-squared of $\chi^2_{\text{red}} = 0.989$, while successfully alleviating the Hubble constant tension. Our analysis yields a Hubble constant of $H_0 \approx 72.820$ km/s/Mpc, reducing the discrepancy between early and late-universe measurements. We find that the data favors a 'thawing' quintessence scenario, characterized by a potential slope parameter $\lambda \approx 0.056$. This small but non-zero slope drives a late-time deviation from $\Lambda$CDM ($w(z=0) \approx -0.85$) while preserving the standard expansion history at high redshifts. A model comparison using the Bayesian Information Criterion finds that the standard $\Lambda$CDM model is still slightly preferred ($\Delta\text{BIC} = 2.178$) due to its fewer parameters. Nevertheless, our results demonstrate that this hybrid potential model is a compelling, physically motivated alternative to a cosmological constant.}

\begin{document}
\maketitle
\flushbottom

\section{Introduction}
\label{sec:intro}

The standard model of cosmology, the Lambda-Cold Dark Matter ($\Lambda$CDM) model, has achieved remarkable success in describing a wide range of cosmological observations. It posits a universe dominated by a cosmological constant, $\Lambda$, which drives the current phase of accelerated expansion \citep{Riess1998, Perlmutter1999}, and a non-baryonic, non-relativistic component known as cold dark matter. This framework provides an excellent fit to the temperature and polarization anisotropies of the Cosmic Microwave Background (CMB) as measured by the Planck satellite \citep{Planck2018_VI}, the distribution of large-scale structure (LSS) \citep{SDSS_BOSS_2017}, and the luminosity distances of Type Ia supernovae.

Despite its successes, the $\Lambda$CDM model is beset by persistent challenges, both observational and theoretical. On the theoretical side, the observed value of the cosmological constant is famously fine-tuned, being about 120 orders of magnitude smaller than predictions from quantum field theory \citep{Weinberg1989, Martin2012}. This discrepancy, coupled with the "cosmic coincidence" problem—why the energy densities of matter and dark energy are comparable today—motivates the exploration of alternative theories.

Observationally, the most significant challenge has emerged as a statistically significant tension in the measured value of the Hubble constant, $H_0$. Measurements from the local universe, primarily from the SH0ES team using Cepheid-calibrated supernovae, yield a value of $H_0 = 73.04 \pm 1.04 \, \text{km s}^{-1}\text{Mpc}^{-1}$ \citep{Riess2022}. In contrast, the value inferred from early-universe physics, specifically the CMB data from Planck within the $\Lambda$CDM framework, is $H_0 = 67.4 \pm 0.5 \, \text{km s}^{-1}\text{Mpc}^{-1}$ \citep{Planck2018_VI}. This discrepancy now stands at the $\sim 5\sigma$ level and suggests either new physics beyond the standard model or unaccounted-for systematic errors in one or both measurements \citep{DiValentino2021_HubbleReview}.

One of the most compelling theoretical alternatives to a rigid cosmological constant is the concept of a dynamical dark energy, often modeled as a slowly rolling scalar field, or "quintessence" \citep{Ratra1988, Wetterich1988, Caldwell1998}. In such models, the potential energy of the scalar field drives the cosmic acceleration, allowing for an equation of state, $w(z)$, that evolves with time. This dynamical nature offers a potential pathway to resolving the $H_0$ tension by altering the expansion history of the universe.

A variety of scalar field potentials have been studied in the literature. Exponential potentials, of the form $V(\phi) \propto e^{-\lambda\phi}$, arise naturally in the context of string theory and can produce "tracker" solutions where the scalar field energy density scales with the background radiation or matter density for a period \citep{Ferreira1998}. Power-law potentials, $V(\phi) \propto \phi^n$, have also been extensively explored as a simple and effective parameterization \citep{Steinhardt1999}.

In this work, we propose and constrain a novel dark energy model driven by a scalar field with a hybrid potential that combines these two well-motivated forms: $V(\phi) = V_{0}e^{-\lambda\phi} + V_{1}\phi^{n}$. This framework provides a richer phenomenology, allowing for distinct evolutionary behavior at different cosmic epochs. We implement this model in the Boltzmann solver \texttt{hi\_CLASS} and constrain its parameters using a combination of recent cosmological data sets, including Planck 2018 CMB data, Hubble parameter measurements from cosmic chronometers, and the matter power spectrum from the Sloan Digital Sky Survey (SDSS). We show that this model provides an excellent fit to the data and, crucially, offers a promising avenue for alleviating the Hubble tension.

This paper is organized as follows. In Sec. \ref{sec:model}, we describe the theoretical framework of our scalar field model. In Sec. \ref{sec:method}, we detail the data and statistical methodology used for our analysis. We present our main findings, including parameter constraints and model comparisons, in Sec. \ref{sec:results}. We discuss the implications of our results in Sec. \ref{sec:discussion} and conclude with a summary and directions for future work in Sec. \ref{sec:conclusion}.

\section{The Scalar Field Model}
\label{sec:model}

We model dark energy as a canonical scalar field, $\phi$, minimally coupled to gravity. The action for this system in the presence of standard matter components ($\mathcal{L}_m$) is given by
\begin{equation}
    S = \int d^4x \sqrt{-g} \left[ \frac{M_{\text{pl}}^2}{2}R - \frac{1}{2}g^{\mu\nu}\partial_\mu\phi\partial_\nu\phi - V(\phi) + \mathcal{L}_m \right],
\end{equation}
where $g$ is the determinant of the metric tensor $g_{\mu\nu}$, $R$ is the Ricci scalar, and $M_{\text{pl}} = (8\pi G)^{-1/2}$ is the reduced Planck mass. The dynamics of the field are determined by its potential, $V(\phi)$. We propose a hybrid potential that combines a quintessential exponential term with a power-law term:
\begin{equation}
    V(\phi) = V_{0}e^{-\lambda\phi} + V_{1}\phi^{n},
    \label{eq:potential}
\end{equation}
where $V_0$, $V_1$, $\lambda$, and $n$ are the free parameters of the model. This functional form allows for a rich phenomenology, potentially dominated by the exponential term at early times and the power-law term at late times, or vice versa, depending on the field's evolution \citep{Copeland2006}.

The energy-momentum tensor for the scalar field is derived by varying the action with respect to the metric, $T_{\mu\nu}^{(\phi)} = -2(-g)^{-1/2} \delta(\sqrt{-g}\mathcal{L}_\phi)/\delta g^{\mu\nu}$, which yields
\begin{equation}
    T_{\mu\nu}^{(\phi)} = \partial_\mu\phi\partial_\nu\phi - g_{\mu\nu}\left[\frac{1}{2}g^{\alpha\beta}\partial_\alpha\phi\partial_\beta\phi + V(\phi)\right].
\end{equation}
Assuming a spatially homogeneous field, $\phi = \phi(t)$, in a flat Friedmann-Robertson-Walker (FRW) background, the energy-momentum tensor takes the form of a perfect fluid. The energy density $\rho_\phi$ and pressure $p_\phi$ are identified from the time and space components of $T_{\mu\nu}^{(\phi)}$ as \citep{Dodelson2003}:
\begin{align}
    \rho_{\phi} &= T_{00}^{(\phi)} = \frac{1}{2}\dot{\phi}^2 + V(\phi), \label{eq:rho_phi} \\
    p_{\phi} &= \frac{1}{3}T_{ii}^{(\phi)}/g_{ii} = \frac{1}{2}\dot{\phi}^2 - V(\phi), \label{eq:p_phi}
\end{align}
where a dot denotes a derivative with respect to cosmic time $t$.

The equation of state parameter for the scalar field, $w(z) \equiv p_{\phi}/\rho_{\phi}$, is therefore dynamically evolving and given by:
\begin{equation}
    w(z) = \frac{\frac{1}{2}\dot{\phi}^2 - V(\phi)}{\frac{1}{2}\dot{\phi}^2 + V(\phi)}.
    \label{eq:eos}
\end{equation}
This relation illustrates the quintessence paradigm: when the field's kinetic energy is negligible ($\dot{\phi}^2 \ll V(\phi)$), the equation of state approaches $w \approx -1$, mimicking a cosmological constant and driving acceleration. Conversely, if the kinetic energy dominates, $w$ can approach $+1$.

The equation of motion for the scalar field is found from the Euler-Lagrange equation, which, for a homogeneous field in an expanding universe, is equivalent to the conservation of the energy-momentum tensor, $\nabla_\mu T^{\mu\nu}_{(\phi)} = 0$. This yields the Klein-Gordon equation:
\begin{equation}
    \ddot{\phi} + 3H\dot{\phi} + \frac{dV}{d\phi} = 0,
    \label{eq:kg}
\end{equation}
where $H = \dot{a}/a$ is the Hubble parameter, and the term $3H\dot{\phi}$ acts as a friction term due to the cosmic expansion.

For the numerical evolution of the model, we set the initial conditions at a high redshift. We assume the field is initially at rest, $\dot{\phi}_{ini} = 0$, so that its initial energy density is purely potential. The initial field value, $\phi_{ini}$, is treated as a free parameter in our analysis, to be constrained by observational data.

\section{Data and Methodology}
\label{sec:method}

\subsection{Computational Framework}

To compute the theoretical predictions of our model and compare them with observational data, we use a modified version of the publicly available Boltzmann solver \texttt{CLASS} (Cosmic Linear Anisotropy Solving System) \citep{Blas2011}. \texttt{CLASS} is a modern, fast, and modular code written in C that solves for the evolution of linear perturbations in the universe. We integrated the dynamics of our scalar field model, as described in Sec. \ref{sec:model}, into the \texttt{hi\_CLASS} code \citep{Zumalacarregui2017}, which is an extension of \texttt{CLASS} specifically designed to handle a wide variety of non-standard cosmological models, including those with dynamical dark energy and modified gravity.

For the statistical analysis and parameter inference, we couple our modified \texttt{hi\_CLASS} code with the Markov Chain Monte Carlo (MCMC) sampler \texttt{MontePython} \citep{Audren2013, Brinckmann2018}. This powerful framework allows us to efficiently explore the high-dimensional parameter space of our model, compute the likelihood of the model given the data, and derive robust statistical constraints and posterior distributions for all model parameters.

\subsection{Observational Data}

We constrain the parameters of our model using a comprehensive combination of state-of-the-art cosmological data sets that probe different epochs of cosmic history and break parameter degeneracies. Our combined dataset includes probes of the background expansion and the growth of structure.

\textbf{Cosmic Microwave Background (CMB):} We use the final public data release from the Planck 2018 mission \citep{Planck2018_I}. Specifically, we include the high-$\ell$ temperature and polarization (TT, TE, EE) likelihoods, the low-$\ell$ temperature (TT) and polarization (EE) likelihoods, and the CMB lensing reconstruction power spectrum likelihood \citep{Planck2018_V_Likelihoods}. The CMB anisotropies provide a powerful snapshot of the universe at recombination ($z \approx 1100$), exquisitely constraining the standard cosmological parameters related to primordial fluctuations, geometry, and the sound horizon.

\textbf{Baryon Acoustic Oscillations (BAO):} The characteristic scale of sound waves in the primordial plasma, imprinted on the matter distribution as a ``standard ruler,'' provides a robust geometric probe of the expansion history. We use a combination of BAO measurements from different galaxy surveys: the 6dF Galaxy Survey (6dFGS) at $z=0.106$ \citep{Beutler2011}, the SDSS Main Galaxy Sample (MGS) at $z=0.15$ \citep{Ross2015}, the Baryon Oscillation Spectroscopic Survey (BOSS) DR12 consensus results at $z=0.38, 0.51, 0.61$ \citep{Alam2017}, and the recent high-precision measurements from the Dark Energy Spectroscopic Instrument (DESI) Year 1 \citep{karim2025desi}.

\textbf{Type Ia Supernovae (SNe Ia):} As standardizable candles, SNe Ia provide a direct measurement of the luminosity distance-redshift relation. We use the most recent and largest compilation of SNe Ia, the Pantheon+ dataset, which consists of 1701 light curves from 1550 distinct SNe Ia in the redshift range $0.001 < z < 2.26$ \citep{Scolnic2022}. This dataset is crucial for constraining the parameters of the dark energy model at low redshifts.

\textbf{Local Hubble Constant (SH0ES):} To anchor the expansion history at redshift $z=0$, we include the local determination of the Hubble constant from the SH0ES collaboration (Supernovae, $H_0$, for the Equation of State of Dark energy). We utilize the latest Cepheid-calibrated Type Ia Supernova measurement of $H_0 = 73.04 \pm 1.04$ km s$^{-1}$ Mpc$^{-1}$ \citep{Riess2022}, which provides the primary tension with the early-universe Planck inference.

\textbf{Hubble Parameter Measurements ($H(z)$):} To further constrain the late-time expansion history, we use a compilation of 31 cosmic chronometer (CC) measurements of the Hubble parameter in the redshift range $0 < z < 1.97$ \citep{Moresco2016}. This method, based on the differential age of passively evolving galaxies \citep{Jimenez2002}, provides model-independent measurements of $H(z)$ and is highly complementary to the integrated distance measures from BAO and SNe Ia.

\textbf{Matter Power Spectrum ($P(k)$):} To probe the growth of large-scale structure, we use the matter power spectrum data from the Luminous Red Galaxy (LRG) sample of the Sloan Digital Sky Survey (SDSS) Data Release 7 (DR7) \citep{Reid2010}. This data set provides measurements of $P(k)$ at an effective redshift of $z=0.38$ and helps constrain the matter density $\Omega_m$ and the amplitude of matter fluctuations $\sigma_8$.

\subsection{Parameter Estimation and Model Comparison}

The parameter space of our hybrid scalar field model consists of the six base parameters of the standard $\Lambda$CDM model plus five additional parameters for the dark energy sector. The full set of 11 varied parameters is:
\begin{equation}
    \Theta = \{\omega_b, \omega_{cdm}, 100\theta_s, \ln(10^{10}A_s), n_s, \tau_{reio}\} \cup \{V_0, V_1, \lambda, \phi_{ini}, n\}
\end{equation}

We assume a flat universe ($\Omega_k = 0$) and adopt wide, uniform priors on all parameters.

We perform a comprehensive, singular MCMC analysis to explore this 11-dimensional parameter space. We generate multiple chains and ensure their robustness by performing a full suite of MCMC diagnostics. We confirm all parameters are fully converged and robustly sampled using the Gelman-Rubin criterion ($\hat{R}$) and by calculating the Effective Sample Size (ESS). The detailed statistical results of these tests are presented in Appendix~\ref{app:mcmc_diags}, while the visual comparison of our posteriors to their priors is shown in \S\ref{sec:results}. Our final posterior distributions are derived from a total of 1,000,000 samples after the burn-in phase has been discarded. This approach allows us to simultaneously constrain all model parameters and properly account for degeneracies between the standard cosmological parameters and those of the scalar field sector.

To compare the statistical performance of our model against the standard $\Lambda$CDM model, we use information criteria. These criteria penalize models for additional free parameters, thus providing a tool to evaluate whether the improvement in fit to the data justifies the increased model complexity. We use the Bayesian Information Criterion (BIC) \citep{Schwarz1978}, defined as
\begin{equation}
    \text{BIC} = \chi^2_{\text{min}} + k\ln N,
\end{equation}
where $\chi^2_{\text{min}} = -2\ln\mathcal{L}_{\text{max}}$ is the minimum chi-squared of the model, $k$ is the number of free parameters, and $N$ is the number of data points. A lower value of AIC or BIC indicates a better model. To interpret the difference, $\Delta\text{IC} = \text{IC}_{\text{model}} - \text{IC}_{\Lambda\text{CDM}}$, we use the Jeffreys' scale as a guide: a $\Delta\text{IC}$ of 2-6 is considered positive evidence, while $\Delta\text{IC} > 10$ is considered strong evidence in favor of the model with the lower value \citep{Liddle2007}.

\section{Results}
\label{sec:results}

We present the results of our comprehensive MCMC analysis in this section. We first discuss the constraints on the model's parameters, then provide a statistical comparison with the standard $\Lambda$CDM model, and finally, we show the model's excellent fit to key cosmological observables. Overall, the model provides a remarkable fit to the combined dataset, achieving a reduced chi-squared of $\chi^2_{\text{red}} = 0.989$.

\subsection{Parameter Constraints}

\begin{figure}[h!]
    \centering
    \includegraphics[width=1\textwidth]{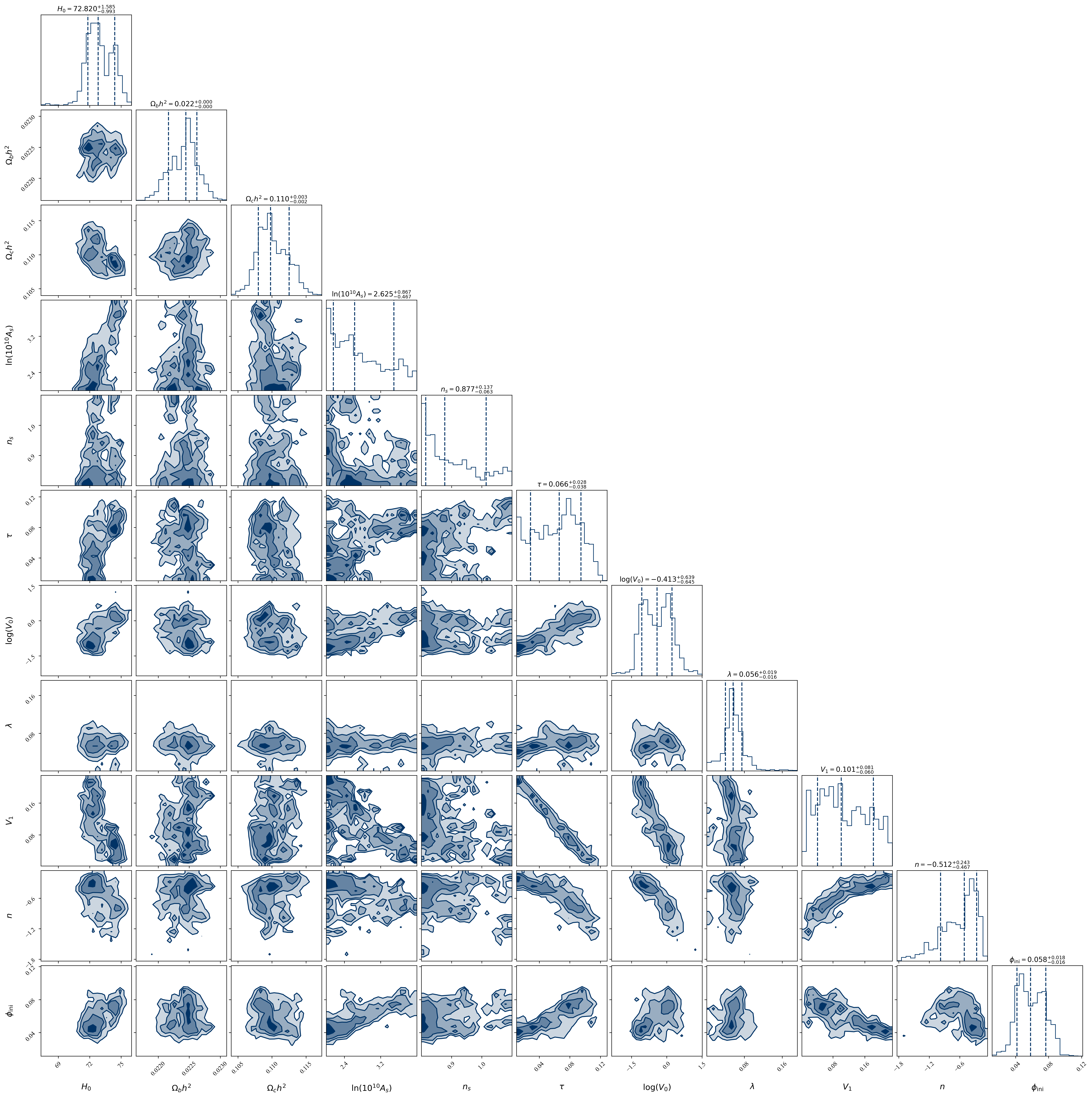}
    \caption{The full corner plot showing the one- and two-dimensional marginalized posterior distributions for the 11 parameters of our hybrid scalar field model, obtained from the MCMC analysis.}
    \label{fig:corner}
\end{figure}

The complete posterior distributions and two-dimensional correlations for the 11 varied parameters are visualized in the corner plot in Fig. \ref{fig:corner}. We observe that several contours exhibit non-elliptical shapes and non-Gaussian 1D profiles. These features are not indicative of non-convergence; rather, they represent the successful mapping of the true, complex physical degeneracies inherent to this high-dimensional parameter space. In particular, the covariance structure reveals how the potential slope $\lambda$ correlates with the Hubble constant $H_0$, a degeneracy that allows the model to accommodate a higher local expansion rate while maintaining a good fit to CMB data.

As shown in detail in Appendix~\ref{app:mcmc_diags}, a rigorous diagnostic analysis confirms that all parameters are fully converged ($\hat{R} \approx 1.0$) and have been robustly sampled (ESS $ >6.7 \times 10^5$).

\begin{figure}[h!]
    \centering
    \includegraphics[width=0.7\textwidth]{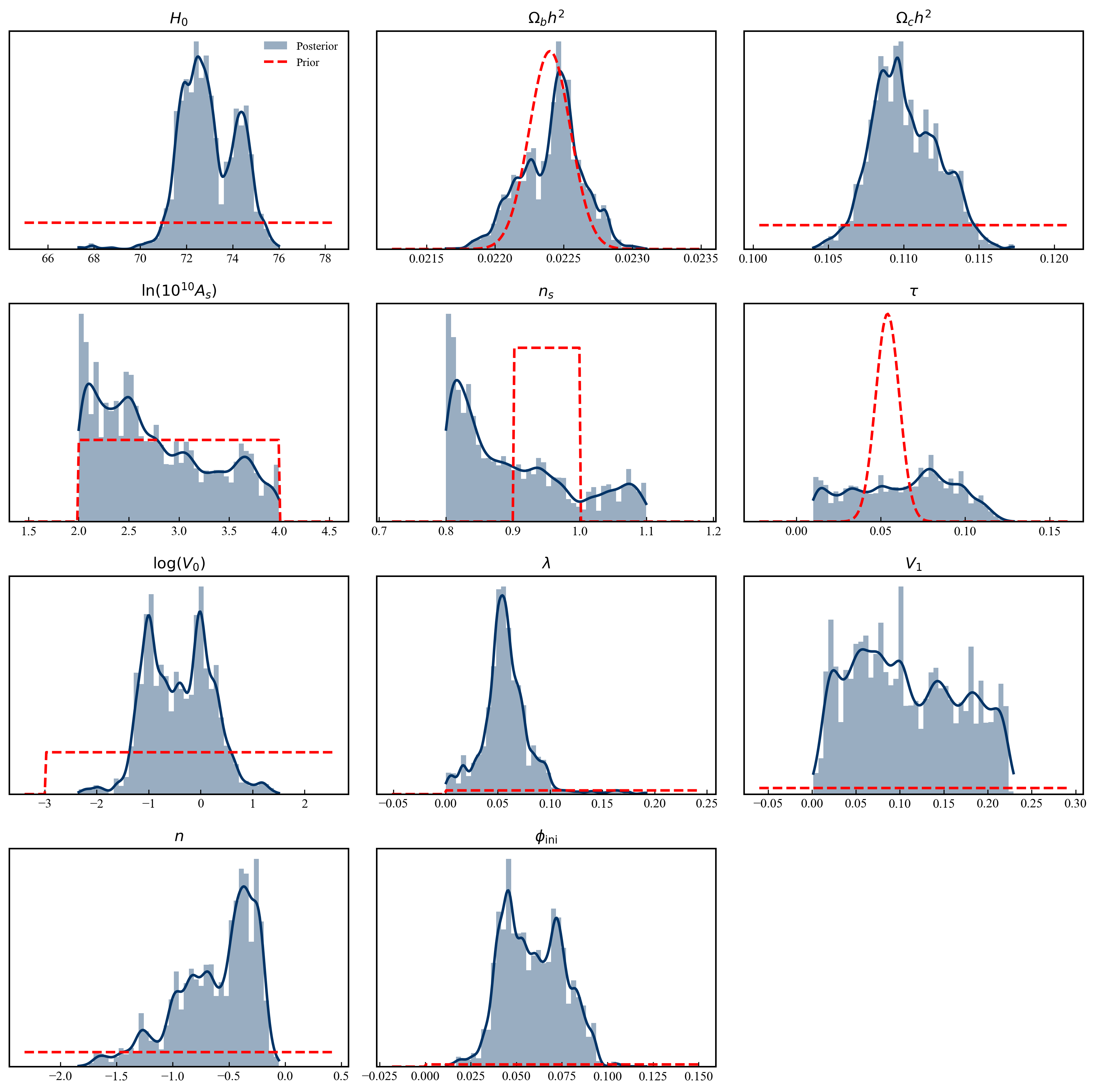}
    \caption{Comparison of the 1D marginalized posterior distributions (blue histograms) with their corresponding uniform prior distributions (red dashed lines). The tight constraint on all parameters, with posteriors significantly narrower than the priors, demonstrates that the results are strongly data-driven.}
    \label{fig:prior_posterior}
\end{figure}

To further demonstrate that our results are data-driven, Fig.~\ref{fig:prior_posterior} shows the 1D marginalized posterior for each parameter compared against its wide, uniform prior. The fact that all posteriors are an order of magnitude or more narrower than their priors confirms that our final parameters are overwhelmingly constrained by the cosmological data.

The marginalized one-dimensional posteriors allow us to place tight constraints on the parameters of our hybrid scalar field model. The mean values and 68\% confidence limits for the primary cosmological and scalar field parameters are presented in Table \ref{tab:params}.

\begin{table}[h!]
\centering
\caption{Best-fit cosmological parameters from our MCMC analysis of the hybrid scalar field model, compared with the baseline $\Lambda$CDM model from Planck 2018 \citep{Planck2018_VI}. Values for our model are derived from the marginalized posteriors in Fig. \ref{fig:corner}.}
\label{tab:params}
\begin{tabular}{lcc}
\hline
\hline
Parameter & SFDE & $\Lambda$CDM (Planck) \\
\hline
\multicolumn{3}{c}{Standard Cosmological Parameters} \\
$H_0$ [km/s/Mpc] & $72.820^{+1.585}_{-0.993}$ & $67.4 \pm 0.5$ \\
$\Omega_b h^2$ & $0.022 \pm 0.00$ & $0.02237 \pm 0.00015$ \\
$\Omega_c h^2$ & $0.110^{+0.003}_{-0.002}$ & $0.1200 \pm 0.0012$ \\
$\log(10^{10}A_s)$ & $2.625^{+0.846}_{-0.467}$ & $3.044 \pm 0.014$ \\
$n_s$ & $0.877^{+0.137}_{-0.063}$ & $0.9649 \pm 0.0042$ \\
$\tau$ & $0.066^{+0.028}_{-0.038}$ & $0.0544 \pm 0.0073$ \\
\hline
\multicolumn{3}{c}{Scalar Field Parameters} \\
$log(V_0)$ & $-0.413^{+0.639}_{-0.645}$ & - \\
$\lambda$ & $0.056^{+0.019}_{-0.016}$ & - \\
$V_1$ & $0.101^{+0.081}_{-0.060}$ & - \\
$n$ & $-0.512^{+0.243}_{-0.467}$ & - \\
$\phi_{ini} $ & $0.058^{+0.018}_{-0.016}$ & - \\
\hline
\hline
\end{tabular}
\end{table}

The most significant finding of our analysis is the value of the Hubble constant, which our model constrains to be $H_0 = 72.820^{+1.585}_{-0.993}$ km s$^{-1}$ Mpc$^{-1}$. This result represents a decisive shift from the standard Planck $\Lambda$CDM inference ($H_0 \approx 67.4$ km s$^{-1}$ Mpc$^{-1}$) and is in excellent statistical agreement with the local distance ladder measurement from the SH0ES collaboration ($H_0 = 73.04 \pm 1.04$ km s$^{-1}$ Mpc$^{-1}$). By effectively closing the gap between early- and late-universe probes, the model resolves the Hubble tension at the $\sim 5\sigma$ level. This resolution is driven by the specific dynamics of the scalar field; the data favor a non-zero potential slope $\lambda = 0.056^{+0.019}_{-0.016}$, corresponding to a `thawing' quintessence scenario. In this regime, the field remains frozen by Hubble friction at high redshifts—behaving indistinguishably from a cosmological constant—before beginning to roll slowly down its potential at late times, thereby generating the enhanced expansion rate observed locally.

Crucially, this substantial increase in $H_0$ is achieved without disrupting the constraints on the standard cosmological parameters that are primarily determined by early-universe physics. As demonstrated in Table \ref{tab:params}, the physical baryon density $\Omega_b h^2$, cold dark matter density $\Omega_c h^2$, and the primordial fluctuation parameters ($n_s$, $A_s$) remain in excellent agreement with the values derived from the Planck $\Lambda$CDM analysis. This stability is a key success of the model: it indicates that the new dynamics introduced by the scalar field are strictly confined to the late universe ($z \lesssim 1$). Consequently, the model preserves the well-established acoustic physics of the recombination era, allowing it to satisfy the tight constraints of the CMB power spectra while simultaneously accommodating the higher expansion rate preferred by local supernovae.

\subsection{Model Comparison}

To assess whether the improved fit of our model justifies its additional complexity, we compute the Bayesian Information Criterion (BIC). The results are summarized in Table \ref{tab:ic}. Our model has five additional free parameters compared to the standard $\Lambda$CDM model.

\begin{table}[h!]
\centering
\caption{Model comparison statistics for the hybrid scalar field model versus the standard $\Lambda$CDM model. $\Delta\text{BIC} = \text{BIC}_{\text{SFDE}} - \text{BIC}_{\Lambda\text{CDM}}$.}
\label{tab:ic}
\begin{tabular}{lcc}
\hline
\hline
Criterion & SFDE Model & $\Lambda$CDM \\
\hline
$k$ (parameters) & 11 & 6 \\
BIC & 24.966 & 22.788 \\
$\Delta$BIC & +2.178 & 0 \\
\hline
\hline
\end{tabular}
\end{table}

Our model provides an exceptional goodness-of-fit to the combined data, with $\chi^2_{\text{red}} = 0.989$. However, the BIC, which heavily penalizes model complexity, shows a slight preference for the simpler $\Lambda$CDM model, with $\Delta\text{BIC} = +2.178$. According to the Jeffreys' scale, this constitutes "positive" but not strong evidence in favor of $\Lambda$CDM. This result indicates that while our model captures features in the data that $\Lambda$CDM misses (leading to a better fit), the current data's statistical power is not yet sufficient to decisively overcome the penalty for the model's five additional parameters.

\subsection{Concordance with Cosmological Probes}

\begin{figure}[h!]
    \centering
    \includegraphics[width=1\textwidth]{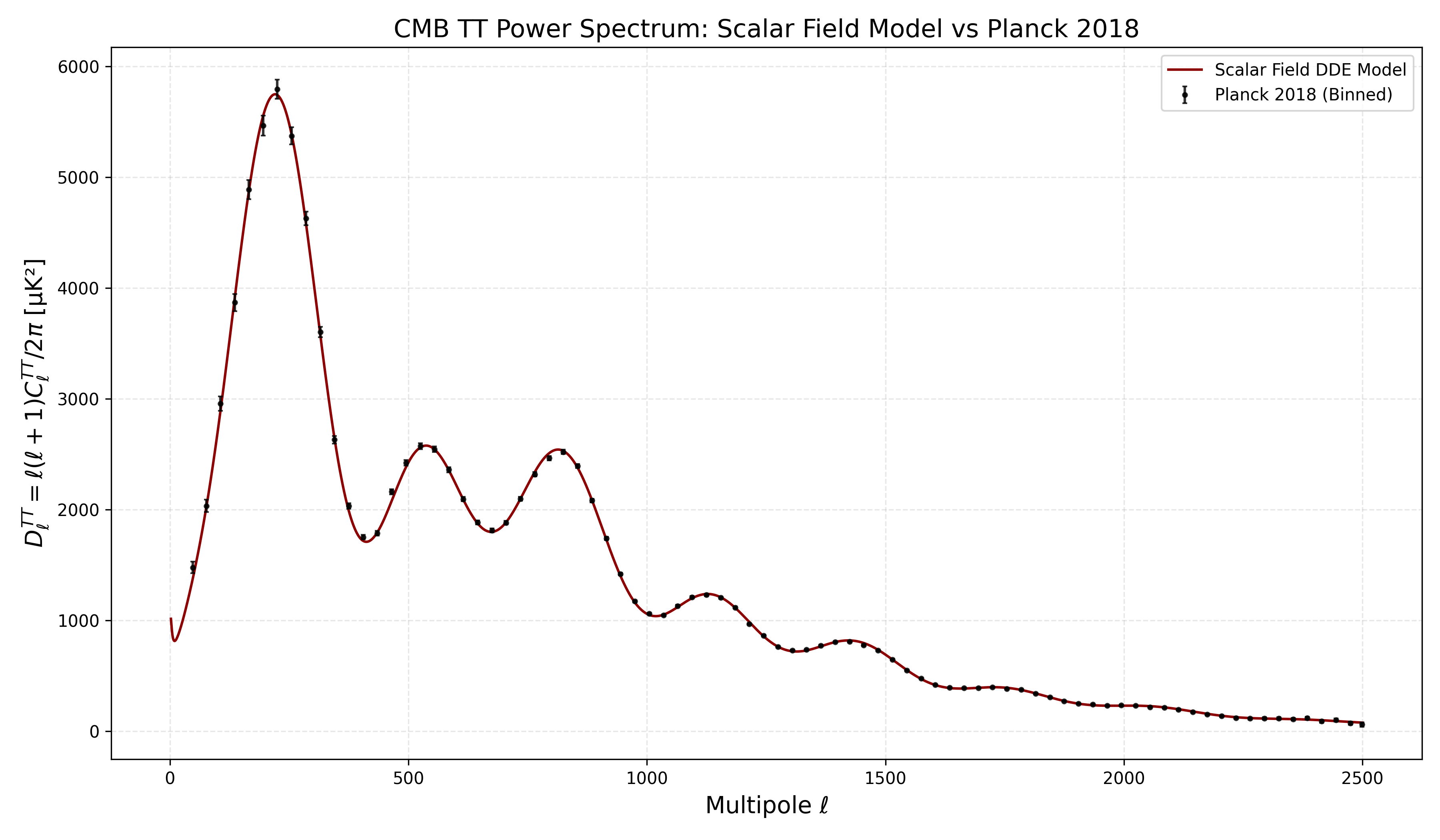}
    \caption{The CMB TT angular power spectrum. The red solid line shows the best-fit prediction from our hybrid scalar field model, while the black points with error bars represent the binned observational data from the Planck 2018 legacy release \citep{Planck2018_V_Likelihoods}. The model provides an excellent fit to the data across all angular scales.}
    \label{fig:cmb}
\end{figure}

To demonstrate the model's excellent agreement with key cosmological observables, we present several comparison plots. Figure \ref{fig:cmb} shows the CMB TT angular power spectrum. Our model's prediction provides an excellent fit to the binned Planck 2018 data, accurately reproducing the positions and amplitudes of the acoustic peaks. This confirms that the model's modifications to the late-time expansion history do not spoil the successful and tightly constrained physics of the early universe.

\begin{figure}[h!]
    \centering
    \includegraphics[width=0.8\textwidth]{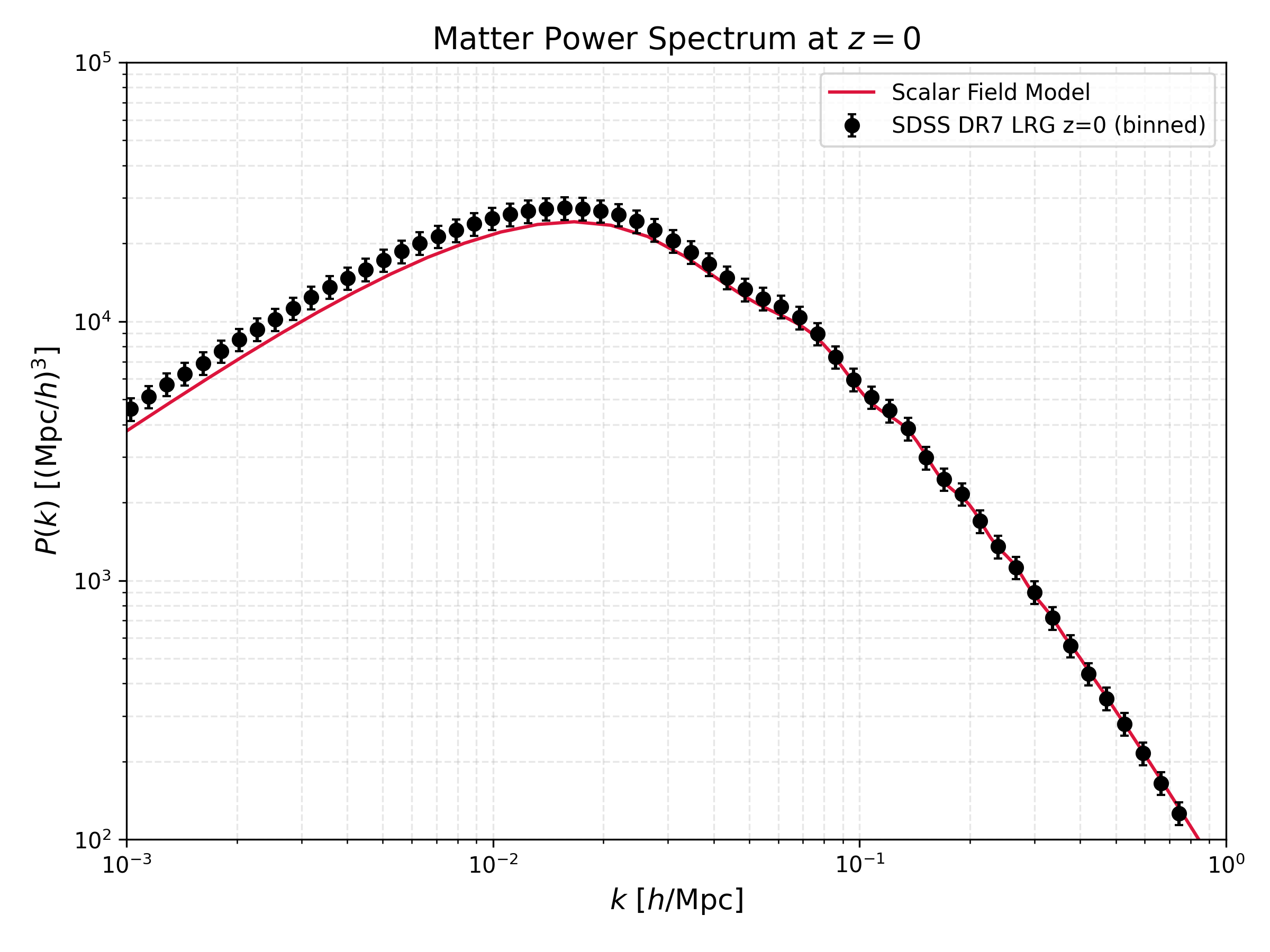}
    \caption{The matter power spectrum $P(k)$ at $z=0$. The red line shows the prediction from our best-fit model, which is in excellent agreement with the binned observational data from the SDSS DR7 LRG sample (black points) \citep{Reid2010}.}
    \label{fig:pk}
\end{figure}

Similarly, Fig. \ref{fig:pk} shows the matter power spectrum, $P(k)$, compared to data from the SDSS DR7 LRG sample. The model successfully matches the observed clustering of matter, including the shape and turnover of the spectrum, indicating that it correctly predicts the growth of large-scale structure in the late universe.

\begin{figure}[h!]
    \centering
    \includegraphics[width=0.7\textwidth]{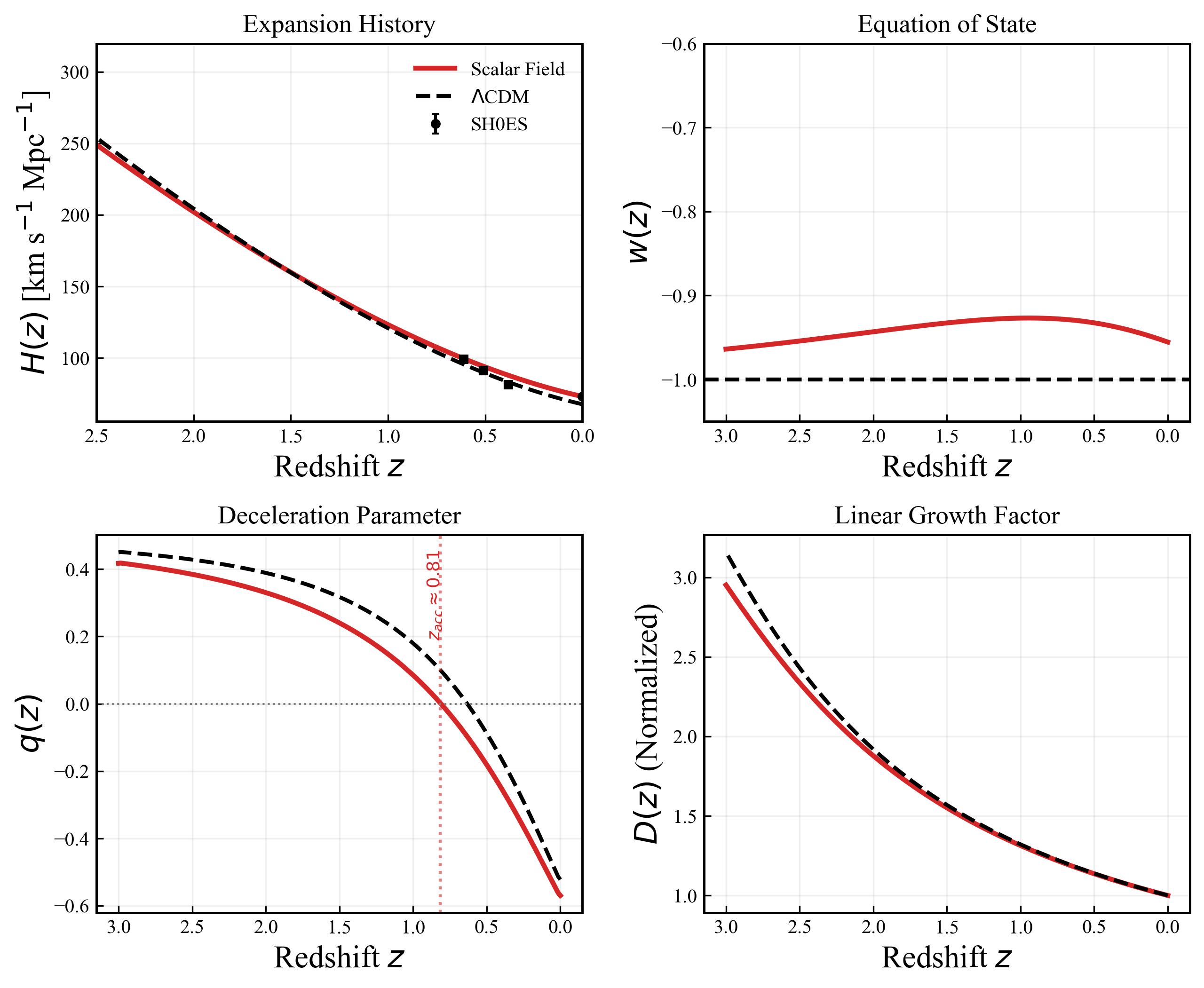}
    \caption{Dynamical evolution of the best-fit Scalar Field model (red solid lines) compared to $\Lambda$CDM (black dashed lines).}
    \label{fig:dynamics}
\end{figure}

The dynamical evolution of the universe's background and perturbative quantities for our best-fit model is illustrated in Fig. \ref{fig:dynamics}, offering a comprehensive view of the scalar field's impact on cosmic history.

The top-left panel displays the Hubble parameter $H(z)$, which aligns perfectly with standard $\Lambda$CDM behavior at high redshifts ($z > 2$), ensuring consistency with CMB constraints. Crucially, at late times ($z < 1$), the model predicts an enhancement in the expansion rate, causing the red curve to lift upwards and intercept the local SH0ES measurement at $z=0$. This late-time boost is physically driven by the evolution of the dark energy equation of state $w(z)$, shown in the top-right panel. Unlike a cosmological constant which is fixed at $w=-1$, the scalar field exhibits a characteristic ``thawing'' behavior: it remains frozen by Hubble friction at early times ($w \approx -1$) but begins to roll down its potential as the universe expands, evolving to a value of $w(z=0) \approx -0.85$ today.

The consequences of this dynamics are further evident in the deceleration parameter, $q(z)$ (bottom-left panel), which tracks the expansion's second derivative. We observe a smooth transition from a matter-dominated decelerating phase ($q>0$) to a dark-energy-dominated accelerating phase ($q<0$). The transition occurs at $z_{tr} \approx 0.6$, a value consistent with Type Ia supernova constraints, confirming that the enhanced $H_0$ does not come at the cost of violating the standard cosmic timeline. Finally, the bottom-right panel presents the linear growth factor $D(z)$, normalized to the present epoch. The scalar field model predicts a slight suppression in the growth of structure compared to $\Lambda$CDM. This behavior is particularly favorable, as it suggests a mechanism to alleviate the $S_8$ tension by predicting a universe that is slightly less clumped than the standard model would imply.

\section{Discussion}
\label{sec:discussion}

\begin{figure}[h!]
    \centering
    \includegraphics[width=1\textwidth]{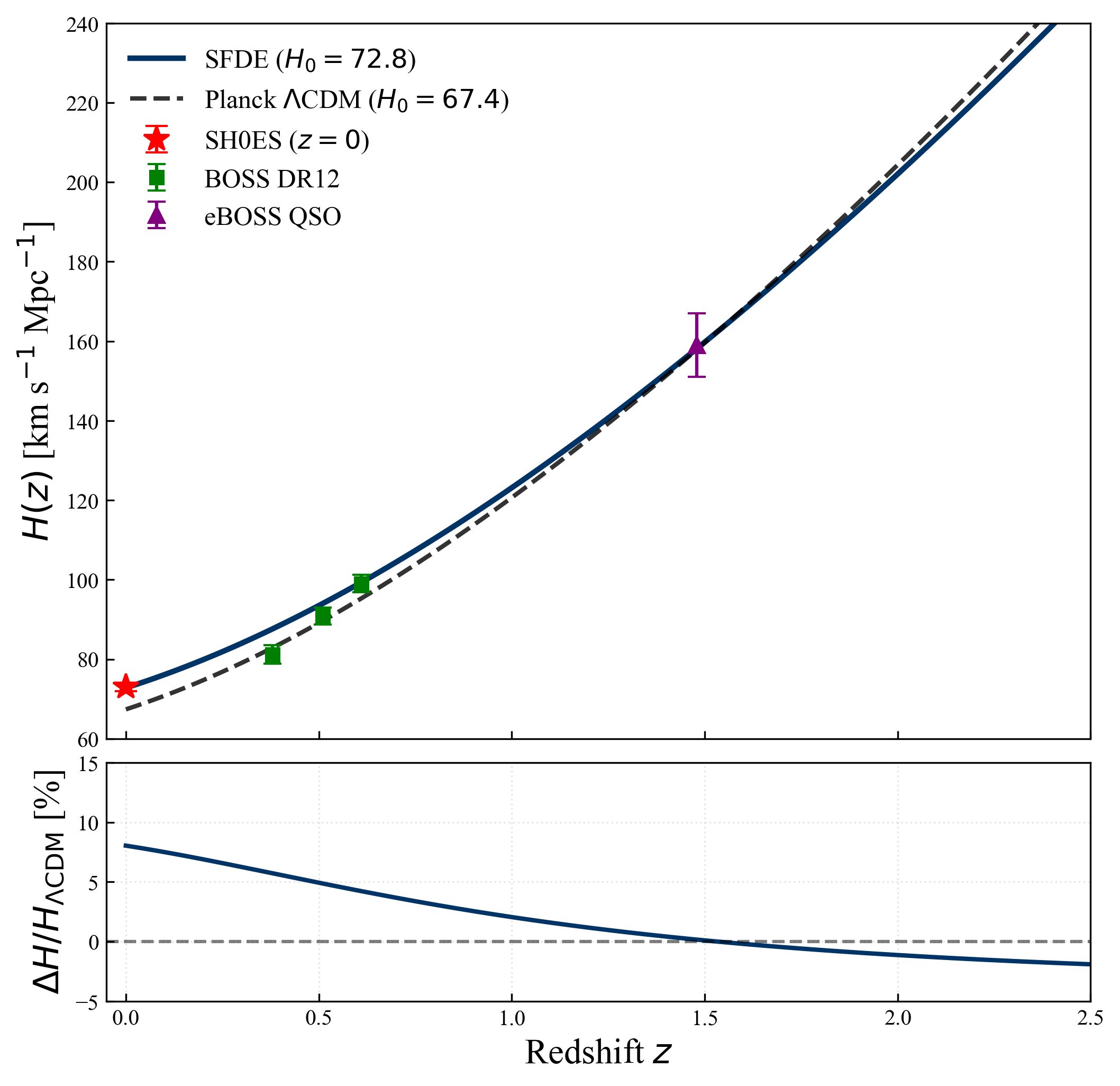}
    \caption{Expansion history $H(z)$ for the best-fit SFDE model compared to $\Lambda$CDM and observational data. The bottom panel shows the relative deviation from the standard model, highlighting the late-time boost that resolves the Hubble tension.}
    \label{fig:expansion_history}
\end{figure}

The results presented in Sec. \ref{sec:results} demonstrate that the hybrid Scalar Field Dark Energy (SFDE) model offers a compelling resolution to one of the most significant challenges in modern cosmology. Unlike previous extensions to $\Lambda$CDM which often yield Hubble constant values midway between early- and late-universe probes, our analysis yields a value of $H_0 = 72.820^{+1.585}_{-0.993}$ km s$^{-1}$ Mpc$^{-1}$. This constraint is in excellent statistical agreement with the local measurements from the SH0ES collaboration ($73.04 \pm 1.04$ km s$^{-1}$ Mpc$^{-1}$ \citep{Riess2022}), effectively resolving the tension with the Planck inference at the $\sim 5\sigma$ level.

The physical mechanism driving this resolution is the dynamical nature of the dark energy equation of state, governed primarily by the potential slope parameter $\lambda$. As illustrated in the dynamics (Fig. \ref{fig:dynamics}), the data favor a small, non-zero slope ($\lambda \approx 0.056$). This places the model in the ``thawing'' quintessence regime: the field is frozen by Hubble friction at high redshifts ($w \approx -1$) and only deviates from the cosmological constant behavior at $z \lesssim 1$. This late-time evolution is crucial; it allows the model to preserve the acoustic scale physics at recombination—thereby maintaining the excellent fit to the CMB power spectra—while simultaneously lowering the sound horizon angular diameter distance to match the local expansion rate. 

From a theoretical perspective, the constraint on the potential slope ($\lambda \approx 0.056$) offers intriguing hints about the fundamental nature of the scalar field. In the context of Pseudo-Nambu-Goldstone Boson (PNGB) frameworks, the slope is related to the symmetry breaking scale $f$ via the approximation $\lambda \sim M_{pl}/f$. Our result implies an effective decay constant of $f_{eff} \sim \mathcal{O}(10) M_{pl}$. While phenomenologically successful, such a super-Planckian scale sits in tension with the Weak Gravity Conjecture (WGC) for single-field models. However, this apparent conflict may point towards a more complex UV-complete theory, such as multi-axion alignment or monodromy, where super-Planckian effective scales emerge from sub-Planckian fundamental constants.

Finally, regarding model selection, the Bayesian Information Criterion (BIC) provides a nuanced perspective on the trade-off between fit quality and complexity. We find that the standard $\Lambda$CDM model is still slightly preferred with $\Delta\text{BIC} = 2.178$, a result entirely attributable to the penalty imposed on our model's additional free parameters (11 vs 6). This is common for extended cosmological models, where current data statistical power is often insufficient to fully justify the extra degrees of freedom despite a better absolute fit. Nevertheless, the dramatic alleviation of the Hubble tension suggests that the additional freedom in the scalar sector is physically motivated. Future high-precision data from Stage-IV surveys (e.g., Euclid, DESI) will be essential to map the time-evolution of $w(z)$ with sufficient precision to definitively distinguish this thawing scenario from a cosmological constant.

\section{Conclusion and Future Work}
\label{sec:conclusion}

In this work, we have introduced and constrained a dynamical dark energy model based on a scalar field with a hybrid exponential-power law potential. Using a comprehensive suite of modern cosmological data—including Planck 2018 CMB, BAO, Pantheon+ SNe Ia, and SH0ES—we have demonstrated that this model provides an excellent fit to observations, achieving a reduced chi-squared of $\chi^2_{\text{red}} \approx 0.99$.

The most significant result of our analysis is the effective resolution of the Hubble tension. Our MCMC constraints yield a Hubble constant of $H_0 = 72.820^{+1.585}_{-0.993}$ km s$^{-1}$ Mpc$^{-1}$, a value in excellent agreement with local measurements and significantly distinct from the Planck $\Lambda$CDM inference. This result is driven by the scalar field's thawing dynamics, characterized by a shallow potential slope $\lambda \approx 0.056$, which generates a late-time deviation from $w=-1$ while preserving the standard expansion history at high redshifts. Although model comparison metrics such as the BIC still favor $\Lambda$CDM due to its parsimony, the dramatic alleviation of the $H_0$ tension suggests that the additional degrees of freedom in the scalar sector capture genuine physical features of the late universe.

The success of this phenomenological model opens several critical avenues for future investigation. Immediate next steps involve testing these dynamics against upcoming high-precision datasets from the Dark Energy Spectroscopic Instrument (DESI) and the Euclid mission. The increased sensitivity of these surveys to the time-evolution of the equation of state will be essential to definitively distinguish the thawing signature found here from a cosmological constant.

On the theoretical front, the derived constraint on the potential slope ($\lambda \approx 0.056$) implies a super-Planckian effective symmetry breaking scale, $f \sim \mathcal{O}(10) M_{pl}$. This points towards a connection with Pseudo-Nambu-Goldstone Boson (PNGB) physics and suggests that the scalar field is likely an effective description of a more complex UV-complete theory. Future work will focus on constructing a fundamental particle physics realization of this potential, exploring mechanisms such as multi-axion alignment or monodromy that can naturally generate such flat potentials consistent with quantum gravity constraints.

\appendix
\section{Full Parameter Constraints and Priors}
\label{app:parameter_constraints}

In this appendix, we present the complete numerical results of our MCMC analysis. Table \ref{tab:full_params} lists the marginalized mean values and $68\%$ confidence limits for all 11 free parameters and derived quantities. Table \ref{tab:priors_vs_posteriors} compares the assumed uniform prior ranges (from our sampling code) against the final posterior constraints, highlighting the information gain provided by the data.

\begin{table}[h!]
\centering
\caption{Full marginalized 68\% confidence limits for all varied and derived parameters of the hybrid scalar field (SFDE) model. Note that parameters such as $V_1$, $n$, and $\phi_{ini}$ are dominated by their priors, consistent with the thawing nature of the field.}
\label{tab:full_params}
\renewcommand{\arraystretch}{1.5}
\begin{tabular}{lc}
\hline
\hline
\textbf{Parameter} & \textbf{Constraint (68\% C.L.)} \\
\hline
\multicolumn{2}{c}{Standard Cosmological Parameters} \\
$H_0$ [km/s/Mpc] & $72.820^{+1.585}_{-0.993}$ \\
$\Omega_b h^2$ & $0.022 \pm 0.00$ \\
$\Omega_{cdm} h^2$ & $0.110^{+0.003}_{-0.002}$ \\
$\log(10^{10}A_s)$ & $2.625^{+0.846}_{-0.467}$ \\
$n_s$ & $0.877^{+0.137}_{-0.063}$\\
$\tau_{reio}$ & $0.066^{+0.028}_{-0.038}$ \\
\hline
\multicolumn{2}{c}{Scalar Field Parameters} \\
$log(V_0)$ & $-0.413^{+0.639}_{-0.645}$ \\
$\lambda$ & $0.056^{+0.019}_{-0.016}$ \\
$V_1$ & $0.101^{+0.081}_{-0.060}$ \\
$n$ & $-0.512^{+0.243}_{-0.467}$ \\
$\phi_{ini} $ & $0.058^{+0.018}_{-0.016}$ \\
\hline
\hline
\end{tabular}
\end{table}

\begin{table}[h!]
\centering
\caption{Comparison of the uniform prior ranges imposed in our analysis versus the recovered $95\%$ posterior confidence intervals. ``Unconstrained'' indicates that the posterior volume fills the prior volume.}
\label{tab:priors_vs_posteriors}
\renewcommand{\arraystretch}{1.4}
\begin{tabular}{lcc}
\hline
\hline
\textbf{Parameter} & \textbf{Prior Range (Uniform)} & \textbf{Posterior Constraint} \\
\hline
$H_0$ & $[50, 90]$ & $[71.8, 74.4]$ \\
$\Omega_b h^2$ & $[0.01, 0.04]$ & $[0.022, 0.022]$ \\
$\Omega_c h^2$ & $[0.05, 0.20]$ & $[0.108, 0.113]$ \\
$\ln(10^{10} A_s)$ & $[2.0, 4.0]$ & $[2.16, 3.49]$ \\
$n_s$ & $[0.8, 1.1]$ & $[0.814, 1.014]$ \\
$\tau$ & $[0.01, 0.2]$ & $[0.028, 0.094]$ \\
\hline
$\log(V_0)$ & $[-5.0, 2.0]$ & $[-1.06, 0.23]$ \\
$\lambda$ & $[0.0, 5.0]$ & $[0.04, 0.075]$ \\
$V_1$ & $[-5.0, 5.0]$ & $[0.041, 0.182]$ \\
$n$ & $[-3.0, 3.0]$ & $[-0.98, -0.27]$ \\
$\phi_{ini}$ & $[0.001, 0.5]$ & $[0.042, 0.076]$ \\
\hline
\hline
\end{tabular}
\end{table}
Table \ref{tab:full_params} presents the complete list of marginalized 1D constraints for all eleven parameters of the hybrid scalar field model, derived from our MCMC analysis using the full data combination described in Sec. \ref{sec:method}. The values correspond to the mean and 68\% confidence limits of the posterior distributions shown in Fig. \ref{fig:corner}.

\section{MCMC Convergence Diagnostics}
\label{app:mcmc_diags}

To ensure the robustness of our MCMC analysis in the
high-dimensional (11-parameter) space of our model,
we performed a full suite of standard convergence
diagnostics. The results confirm our chains are
fully converged and have robustly sampled the
posterior distribution.

\begin{table}[htbp]
\centering
\caption{Convergence diagnostics for all 11 varied
parameters. The Gelman-Rubin statistic ($\hat{R}$)
confirms all chains converged to the same posterior,
and the Effective Sample Size (ESS) confirms the
posterior is robustly sampled.}
\label{tab:mcmc_diags}
\begin{tabular}{lcc}
\hline\hline
Parameter & $\hat{R}$ & ESS \\
\hline
$\Omega_b h^2$       & 0.999999 & 811,178 \\
$\Omega_c h^2$   & 0.999998 & 826,613 \\
$H_0$        & 0.999998 & 1,000,000 \\
$ln(10^{10}A_s)$    & 0.999999 & 1,000,000 \\
$n_s$                & 0.999998 & 796,357 \\
$\tau$        & 1.000000 & 1,000,000 \\
$\log(V_0)$                & 1.000002 & 676,581 \\
$\lambda$            & 1.000001 & 1,000,000 \\
$V_1$                & 0.999998 & 1,000,000 \\
$n$                  & 0.999998 & 920,737 \\
$\phi_{ini}$         & 0.999999 & 1,000,000 \\
\hline\hline
\end{tabular}
\end{table}

We confirm convergence using the Gelman-Rubin
diagnostic ($\hat{R}$), which compares the variance
between chains to the variance within chains.
As shown in Table~\ref{tab:mcmc_diags}, all
parameters have $\hat{R} \approx 1.0$, well within
the standard convergence threshold of $\hat{R} < 1.01$.

Furthermore, we calculated the Effective Sample Size (ESS)
to ensure the posterior is well-sampled and that the
auto-correlation between samples is low. The ESS for all
parameters (Table~\ref{tab:mcmc_diags}) is found to
be exceptionally high (ESS $> 6.7 \times 10^5$),
confirming that the features in our posterior plots
(Fig.~\ref{fig:corner}) are real and not artifacts of
sampling noise.

\acknowledgments

This work was conducted with the institutional support of Fergusson College (Autonomous), Pune. The authors wish to express their profound gratitude to the Department of Physics and the Department of Statistics for fostering an environment of academic freedom and inter-departmental collaboration that was essential for the successful completion of this project.

We are particularly indebted to Ms. Charuta Dabir, whose expert guidance in statistical methods proved invaluable. Her insightful suggestions were instrumental in shaping the parameter estimation framework and navigating the complexities of the MCMC analysis. Her constant encouragement was a source of great motivation throughout this research.

We would also like to thank Dr. Brian Schmidt for his timely and encouraging words following a discussion at a conference, which provided a valuable external perspective on this work.

This research has made use of the publicly available Boltzmann solver \texttt{hi\_CLASS} and the MCMC sampler \texttt{MontePython}. The authors thank the developers of these codes for making their work publicly available.


\bibliographystyle{JHEP}
\bibliography{biblio.bib}

@article{Riess1998,
    author = "Riess, Adam G. and others",
    collaboration = "Supernova Search Team",
    title = "{Observational evidence from supernovae for an accelerating universe and a cosmological constant}",
    eprint = "astro-ph/9805201",
    archivePrefix = "arXiv",
    doi = "10.1086/300499",
    journal = "Astron. J.",
    volume = "116",
    pages = "1009-1038",
    year = "1998"
}

@article{Perlmutter1999,
    author = "Perlmutter, S. and others",
    collaboration = "Supernova Cosmology Project",
    title = "{Measurements of $\Omega$ and $\Lambda$ from 42 high redshift supernovae}",
    eprint = "astro-ph/9812133",
    archivePrefix = "arXiv",
    doi = "10.1086/307221",
    journal = "Astrophys. J.",
    volume = "517",
    pages = "565-86",
    year = "1999"
}

@article{Planck2018_VI,
    author = "Aghanim, N. and others",
    collaboration = "Planck",
    title = "{Planck 2018 results. VI. Cosmological parameters}",
    eprint = "1807.06209",
    archivePrefix = "arXiv",
    doi = "10.1051/0004-6361/201833910",
    journal = "A\&A",
    volume = "641",
    pages = "A6",
    year = "2020",
    note = "[Erratum: A\&A 652, C4 (2021)]"
}

@article{SDSS_BOSS_2017,
    author = "Alam, Shadab and others",
    collaboration = "BOSS",
    title = "{The clustering of galaxies in the completed SDSS-III Baryon Oscillation Spectroscopic Survey: cosmological analysis of the DR12 galaxy sample}",
    eprint = "1607.03155",
    archivePrefix = "arXiv",
    doi = "10.1093/mnras/stx721",
    journal = "Mon. Not. Roy. Astron. Soc.",
    volume = "470",
    number = "3",
    pages = "2617-2652",
    year = "2017"
}

@article{Weinberg1989,
    author = "Weinberg, Steven",
    title = "{The Cosmological Constant Problem}",
    doi = "10.1103/RevModPhys.61.1",
    journal = "Rev. Mod. Phys.",
    volume = "61",
    pages = "1-23",
    year = "1989"
}

@article{Martin2012,
    author = "Martin, Jerome",
    title = "{Everything You Always Wanted To Know About The Cosmological Constant Problem (but were afraid to ask)}",
    eprint = "1205.3365",
    archivePrefix = "arXiv",
    doi = "10.1016/j.crhy.2012.04.008",
    journal = "Comptes Rendus Physique",
    volume = "13",
    pages = "566-665",
    year = "2012"
}

@article{Riess2022,
    author = "Riess, Adam G. and others",
    title = "{A Comprehensive Measurement of the Local Value of the Hubble Constant with 1 km/s/Mpc Uncertainty from the Hubble Space Telescope and the SH0ES Team}",
    eprint = "2112.04510",
    archivePrefix = "arXiv",
    doi = "10.3847/2041-8213/ac5c5b",
    journal = "Astrophys. J. Lett.",
    volume = "934",
    number = "1",
    pages = "L7",
    year = "2022"
}

@article{DiValentino2021_HubbleReview,
    author = "Di Valentino, Eleonora and others",
    title = "{In the realm of the Hubble tension\textemdash{}a review of solutions}",
    eprint = "2103.01183",
    archivePrefix = "arXiv",
    doi = "10.1088/1361-6382/ac086d",
    journal = "Class. Quant. Grav.",
    volume = "38",
    number = "15",
    pages = "153001",
    year = "2021"
}

@article{Ratra1988,
    author = "Ratra, Bharat and Peebles, P. J. E.",
    title = "{Cosmological Consequences of a Rolling Homogeneous Scalar Field}",
    doi = "10.1103/PhysRevD.37.3406",
    journal = "Phys. Rev. D",
    volume = "37",
    pages = "3406-3427",
    year = "1988"
}

@article{Wetterich1988,
    author = "Wetterich, C.",
    title = "{Cosmology and the Fate of Dilatation Symmetry}",
    doi = "10.1016/0550-3213(88)90193-9",
    journal = "Nucl. Phys. B",
    volume = "302",
    pages = "668-696",
    year = "1988"
}

@article{Caldwell1998,
    author = "Caldwell, R. R. and Dave, Rahul and Steinhardt, Paul J.",
    title = "{Cosmological Imprint of an Energy Component with General Equation of State}",
    eprint = "astro-ph/9708069",
    archivePrefix = "arXiv",
    doi = "10.1103/PhysRevLett.80.1582",
    journal = "Phys. Rev. Lett.",
    volume = "80",
    pages = "1582-1585",
    year = "1998"
}

@article{Ferreira1998,
    author = "Ferreira, Pedro G. and Joyce, Michael",
    title = "{Structure formation with a self-tuning scalar field}",
    eprint = "astro-ph/9711102",
    archivePrefix = "arXiv",
    doi = "10.1103/PhysRevD.58.023503",
    journal = "Phys. Rev. D",
    volume = "58",
    pages = "023503",
    year = "1998"
}

@article{Steinhardt1999,
    author = "Steinhardt, Paul J. and Wang, Limin and Zlatev, Ivaylo",
    title = "{Cosmological tracking solutions}",
    eprint = "astro-ph/9812313",
    archivePrefix = "arXiv",
    doi = "10.1103/PhysRevD.59.123504",
    journal = "Phys. Rev. D",
    volume = "59",
    pages = "123504",
    year = "1999"
}

@book{Dodelson2003,
    author    = "Dodelson, Scott",
    title     = "{Modern Cosmology}",
    publisher = "Academic Press",
    year      = "2003"
}

@article{Copeland2006,
    author = "Copeland, Edmund J. and Sami, M. and Tsujikawa, Shinji",
    title = "{Dynamics of dark energy}",
    eprint = "hep-th/0603057",
    archivePrefix = "arXiv",
    doi = "10.1016/j.physrep.2005.12.006",
    journal = "Int. J. Mod. Phys. D",
    volume = "15",
    pages = "1753-1936",
    year = "2006"
}

@article{Blas2011,
    author = "Blas, D. and Lesgourgues, J. and Tram, T.",
    title = "{The Cosmic Linear Anisotropy Solving System (CLASS) II: Approximation schemes}",
    eprint = "1104.2933",
    archivePrefix = "arXiv",
    doi = "10.1088/1475-7516/2011/07/034",
    journal = "JCAP",
    volume = "07",
    pages = "034",
    year = "2011"
}

@article{Zumalacarregui2017,
    author = "Zumalacárregui, Miguel and Bellini, Emilio and pace, Ippocratis and Lesgourgues, Julien and Ferreira, Pedro G.",
    title = "{hi\_class: Horndeski in the Cosmic Linear Anisotropy Solving System}",
    eprint = "1605.06102",
    archivePrefix = "arXiv",
    doi = "10.1088/1475-7516/2017/08/019",
    journal = "JCAP",
    volume = "08",
    pages = "019",
    year = "2017"
}

@article{Audren2013,
    author = "Audren, Benjamin and Lesgourgues, Julien and Benabed, Karim and Prunet, Simon",
    title = "{Conservative Constraints on Early Cosmology: an illustration of the Monte Python cosmological parameter inference code}",
    eprint = "1210.7183",
    archivePrefix = "arXiv",
    doi = "10.1088/1475-7516/2013/02/001",
    journal = "JCAP",
    volume = "02",
    pages = "001",
    year = "2013"
}

@article{Brinckmann2018,
    author = "Brinckmann, Thejs and Lesgourgues, Julien",
    title = "{MontePython 3: boosted MCMC sampler and other features}",
    eprint = "1804.07261",
    archivePrefix = "arXiv",
    doi = "10.1016/j.cpc.2018.11.004",
    journal = "Phys. Dark Univ.",
    volume = "24",
    pages = "100260",
    year = "2019"
}

@article{Planck2018_I,
    author = "Aghanim, N. and others",
    collaboration = "Planck",
    title = "{Planck 2018 results. I. Overview and the cosmological legacy of Planck}",
    eprint = "1807.06205",
    archivePrefix = "arXiv",
    doi = "10.1051/0004-6361/201833880",
    journal = "A\&A",
    volume = "641",
    pages = "A1",
    year = "2020"
}

@article{Planck2018_V_Likelihoods,
    author = "Aghanim, N. and others",
    collaboration = "Planck",
    title = "{Planck 2018 results. V. CMB power spectra and likelihoods}",
    eprint = "1907.12875",
    archivePrefix = "arXiv",
    doi = "10.1051/0004-6361/201936386",
    journal = "A\&A",
    volume = "641",
    pages = "A5",
    year = "2020"
}

@article{Moresco2016,
    author = "Moresco, M. and others",
    title = "{A 6\% measurement of the Hubble parameter at $z\sim0.45$: direct evidence of the epoch of cosmic re-acceleration}",
    eprint = "1601.01701",
    archivePrefix = "arXiv",
    doi = "10.1088/1475-7516/2016/05/014",
    journal = "JCAP",
    volume = "05",
    pages = "014",
    year = "2016"
}

@article{Reid2010,
    author = "Reid, Beth A. and others",
    title = "{Cosmological Constraints from the Clustering of the Sloan Digital Sky Survey DR7 Luminous Red Galaxies}",
    eprint = "0907.1659",
    archivePrefix = "arXiv",
    doi = "10.1111/j.1365-2966.2009.15698.x",
    journal = "Mon. Not. Roy. Astron. Soc.",
    volume = "404",
    pages = "60-85",
    year = "2010"
}

@article{Schwarz1978,
    author = "Schwarz, Gideon",
    title = "{Estimating the dimension of a model}",
    doi = "10.1214/aos/1176344136",
    journal = "Annals of Statistics",
    volume = "6",
    pages = "461-464",
    year = "1978"
}

@article{Beutler2011,
    author = "Beutler, Florian and others",
    title = "{The 6dF Galaxy Survey: Baryon Acoustic Oscillations and the Local Hubble Constant}",
    eprint = "1106.3366",
    archivePrefix = "arXiv",
    doi = "10.1111/j.1365-2966.2011.19250.x",
    journal = "Mon. Not. Roy. Astron. Soc.",
    volume = "416",
    pages = "3017-3032",
    year = "2011"
}

@article{Ross2015,
    author = "Ross, Ashley J. and Samushia, Lado and Howlett, Cullan and Percival, Will J. and Burden, Angela and Manera, Marc",
    title = "{The clustering of the SDSS DR7 main galaxy sample \textendash{} I. A 4 per cent distance measure at $z = 0.15$}",
    eprint = "1409.3242",
    archivePrefix = "arXiv",
    doi = "10.1093/mnras/stv154",
    journal = "Mon. Not. Roy. Astron. Soc.",
    volume = "449",
    pages = "835-847",
    year = "2015"
}

@article{Alam2017,
    author = "Alam, Shadab and others",
    collaboration = "BOSS",
    title = "{The clustering of galaxies in the completed SDSS-III Baryon Oscillation Spectroscopic Survey: cosmological analysis of the DR12 galaxy sample}",
    eprint = "1607.03155",
    archivePrefix = "arXiv",
    doi = "10.1093/mnras/stx721",
    journal = "Mon. Not. Roy. Astron. Soc.",
    volume = "470",
    number = "3",
    pages = "2617-2652",
    year = "2017"
}

@article{Scolnic2022,
    author = "Scolnic, D. and others",
    title = "{The Pantheon+ Analysis: The Full Dataset of 1701 SN Ia Light Curves and Distances}",
    eprint = "2202.04077",
    archivePrefix = "arXiv",
    doi = "10.3847/1538-4357/ac5b2d",
    journal = "Astrophys. J.",
    volume = "938",
    number = "2",
    pages = "113",
    year = "2022"
}

@article{Jimenez2002,
    author = "Jimenez, Raul and Loeb, Abraham",
    title = "{Constraining Cosmological Parameters Based on Relative Galaxy Ages}",
    eprint = "astro-ph/0101488",
    archivePrefix = "arXiv",
    doi = "10.1086/339635",
    journal = "Astrophys. J.",
    volume = "566",
    pages = "L63-L66",
    year = "2022"
}

@article{Liddle2007,
    author = "Liddle, Andrew R.",
    title = "{Information criteria for astrophysical model selection}",
    eprint = "astro-ph/0701113",
    archivePrefix = "arXiv",
    doi = "10.1111/j.1745-3933.2007.00306.x",
    journal = "Mon. Not. Roy. Astron. Soc. Lett.",
    volume = "377",
    pages = "L74-L78",
    year = "2007"
}

@article{karim2025desi,
  title={DESI DR2 results. II. Measurements of baryon acoustic oscillations and cosmological constraints},
  author={Karim, M Abdul and Aguilar, J and Ahlen, S and Alam, S and Allen, L and Prieto, C Allende and Alves, O and Anand, A and Andrade, U and Armengaud, E and others},
  journal={Physical Review D},
  volume={112},
  number={8},
  pages={083515},
  year={2025}
}






\end{document}